# Self-assembly of colloidal particles in deformation landscapes of electrically driven layer undulations in cholesteric liquid crystals

Michael C. M. Varney,[1,*] Qiaoxuan Zhang,[1,2] Bohdan Senyuk,[1] and Ivan I. Smalyukh[1,2,3,4,†]

[1]*Department of Physics and Soft Materials Research Center, University of Colorado at Boulder, Boulder, Colorado 80309, USA*

[2]*Materials Science and Engineering Program, University of Colorado at Boulder, Boulder, Colorado 80309, USA*

[3]*Department of Electrical, Computer, and Energy Engineering, University of Colorado at Boulder, Boulder, Colorado 80309, USA*

[4]*Renewable and Sustainable Energy Institute, National Renewable Energy Laboratory and University of Colorado at Boulder, Boulder, Colorado 80309, USA*

[*]Present address: Exnodes Inc., Fremont, California 94538, USA.
[†]Corresponding author: ivan.smalyukh@colorado.edu

We study elastic interactions between colloidal particles and deformation landscapes of undulations in a cholesteric liquid crystal under an electric field applied normal to cholesteric layers. The onset of undulation instability is influenced by the presence of colloidal inclusions and, in turn, layers' undulations mediate the spatial patterning of particle locations. We find that the bending of cholesteric layers around a colloidal particle surface prompts the local nucleation of an undulations lattice at electric fields below the well-defined threshold known for liquid crystals without inclusions, and that the onset of the resulting lattice is locally influenced, both dimensionally and orientationally, by the initial arrangements of colloids defined using laser tweezers. Spherical particles tend to spatially localize in the regions of strong distortions of the cholesteric layers, while colloidal nanowires exhibit an additional preference for multistable alignment offset along various vectors of the undulations lattice. Magnetic rotation of superparamagnetic colloidal particles couples with the locally distorted helical axis and undulating cholesteric layers in a manner that allows for a controlled three-dimensional translation of these particles. These interaction modes lend insight into the physics of liquid crystal structure-colloid elastic interactions, as well as point the way towards guided self-assembly of reconfigurable colloidal composites with potential applications in diffraction optics and photonics.

# I. INTRODUCTION

Liquid crystals (LCs) [1,2], including nematic, cholesteric, smectic, and various more exotic thermodynamic phases, such as blue phases, have recently become popular fluid host media for colloidal particle assembly and alignment [3–25]. Compared to isotropic fluid hosts, LCs bring many new possibilities in designing guided colloidal self-assembly arising from elasticity-mediated anisotropic interactions [3], nanoparticle localization by topological defects [11,19,26], topography [27–29] and patterning [30,31] of confinement, and structural switching enabled by the LC's facile response to external fields [10,32]. However, the focus of these studies of LC colloidal dispersions is primarily on using ground-state phases [11], predesigned molecular fields modulations [30,31], or topological defect arrays to mediate self-assembly. Applied external fields allow for additional control of colloidal particles' orientation and localization [33,34] and their directional transport [35,36] in nematic liquid crystals. However, despite a large number of mesomorphic phases with different symmetries and combinations of partial positional and orientational ordering, the available means of using these structured media in positioning, aligning, and assembling nano- and microsized particles are still rather limited. In particular, one fundamental limitation is related to an inability of controlling periodic large-scale localization of particles in these media in order to reversibly form periodic lattices [1,37] in response to electric fields or other external stimuli.

In this work, we propose to use electrically driven cholesteric layer undulations [1,38–43] to localize, align, and induce patterns of colloidal particle assemblies within a cholesteric LC host medium, which allows switching between different organizations of particles using low voltages without, for example, erasing the ground-state helical structure [32], or employing the predesigned molecular patterns [30,31]. Since the helicoidal structure of these LCs is normally featured along a single direction, cholesterics possess an effective layer periodicity equal to half of a cholesteric pitch $p$ (a distance along which molecules, and the director $\mathbf{n}$ describing their local average orientation, twist by $2\pi$ around a helical axis $\chi$) [1,41,44]. These media have many properties characteristic of lamellar phases [41,44], since cholesteric layers tend to preserve equidistance while easily bending in response to external stimuli and boundary conditions [1]. The morphology of layer configurations that satisfies these constraints is rich and includes dislocations [1,2], disclinations [41], focal conic domains [1], undulations [38–43], etc. The last of these, undulations, is perhaps the most studied type of structural instability of equidistant layers that does not alter topology of the uniform ground state [38–40]. This phenomenon of undulations induced by external stimuli is also commonly referred to as the Helfrich-Hurault effect [38–40]. When an electric field is applied across the cholesteric layers of a material with a positive dielectric anisotropy, the layers (and **n**) tend to reorient to become parallel to the field [38–40,42,43], but fully free rotation is hindered by surface anchoring forces at confining substrates. This competition of electric, surface anchoring, and elastic torques results in periodic patterns of undulating layers, with a periodicity dependent on sample thickness and equilibrium pitch and with an undulations amplitude tunable by an applied voltage above a certain well-defined threshold value [38–40,42,43]. Recent progress in controlling cholesteric layer undulations [43] calls for their practical use in mediating colloidal interactions and self-assembly as this has not been achieved so far.

Using laser tweezers, we manipulate colloidal inclusions in three dimensions and demonstrate that particles embedded into cholesteric layers can locally reduce the threshold for forming periodically distorted layered patterns. On the other hand, colloidal particles also interact with the host LC and periodic distortions of its cholesteric layered structure through a combination of elastic and dielectrophoretic effects. These complex interactions result in the formation of periodic potential energy landscapes capable of localizing colloidal microspheres in well-defined locations within the reconfigurable undulations lattices. Moreover, anisotropic colloids, such as GaN nanowires, additionally exhibit controllable multistable alignment with respect to the square-periodic undulations patterns. We show that layer undulations also impinge on the dynamics of magnetically controlled superparamagnetic microparticles, enabling not only a robust control of equilibrium state patterning of particles but also nonequilibrium kinetic processes involving them. These findings are discussed from the standpoint of designing reconfigurable colloidal composites with potential applications in diffraction optics [43,45], nanophotonics [19], electro-optics [1], and other fields of science and engineering.

## II. MATERIALS AND TECHNIQUES

Our work utilizes a cholesteric LC (CLC) composed of nematic hosts 5CB (4-cyano-4'-pentylbiphenyl from Frinton Laboratories, Inc.) or ZLI-3412 (EM Industries) mixed with a chiral agent [cholesteryl pelargonate (Sigma-Aldrich Chemistry) or CB15 (EM Industries)] to produce, respectively, left- or right-handed chiral nematics with a cholesteric pitch in the range of 5–30μm. Cholesteryl pelargonate was mixed only with 5CB while CB15 was used for mixing with both nematic hosts (Table I). These used nematic LC hosts have a positive dielectric anisotropy (Table I), so that their response to an applied electric field manifests itself in the tendency of the director n to rotate to an orientation along the direction of the applied field [1,38,42]. Imaging of the CLC colloidal samples was performed through a combination of polarizing optical microscopy (POM) and three-photon excitation fluorescence polarizing microscopy (3PEF-PM) assembled on the basis of an Olympus IX81 microscope [46,47]. The 3PEF-PM is capable of operating in both epi-detection and forward detection (transmission) modes, with the epi-detection mode being the principle configuration when performing fully holonomic optical and magnetic manipulations. Three-dimensional fluorescence images of the local **n**(**r**) in used mixtures were obtained using a multiphoton-absorption-based excitation of cyanobiphenyl groups of 5CB and CB15 molecules by laser light at 870 nm from a tunable (680–1080 nm) Ti:sapphire oscillator (140 fs, 80 MHz, Chameleon Ultra-II, Coherent) with the resulting fluorescence signal detected within a spectral range of 387–447 nm by a photomultiplier tube H5784-20 (Hamamatsu) [46,47]. POM textures of undulations and colloidal particles were recorded with a charge-coupled device (CCD) camera (Flea, Point Grey Research, Inc.). Optical manipulations of colloidal particles were realized with a holographic optical trapping (HOT) system [46] operating at a wavelength of $\lambda = 1064$ nm and utilizing the same high numerical aperture (NA) objective (100×, NA = 1.4) as used for conventional POM and nonlinear 3PEF-PM imaging. The optical tweezers can be used for positioning colloidal particles both in the plane of the experimental cells and at different distances from the confining substrates [46].

In order to demonstrate the diverse range of possibilities for controlling LC colloidal systems through the exploration of field-induced undulations, we used a variety of colloidal particles in our experiments. Our magnetic colloidal particles are superparamagnetic beads (SPMBs, Dynabead M450, Invitrogen) with a nominal diameter of 4.5 ± 0.1 μm fabricated using ferromagnetic nanoparticles ($\gamma$ $Fe_2O_3$ and $Fe_3O_4$) of ~8-nm-diameter embedded into a highly cross-linked epoxy at a density of ~$10^5$ nanoparticles per bead [48,49]. Random physical orientation of individual ferrite nanoparticles within the epoxy matrix, coupled with thermal moment flipping in each nanoparticle, creates a zero net magnetic moment within the SPMB [49]. An externally applied magnetic field **H** induces the magnetic moment **m** of a given ferrite nanoparticle to favor one direction, which may not necessarily align collinearly with **H**. Since the ferrite nanoparticle orientations are mechanically coupled to the epoxy matrix of the SPMB microsphere, magnetic moment interactions with the applied magnetic field prompt transient SPMB rotation and alignment of its **m** to eventually point along **H**; this interaction subsequently mechanically aligns the SPMB such that its net magnetic moment locks collinear to **H** and follows its rotation [49,50] during our experiments detailed below. In addition to superparamagnetic beads, we also use melamine resin (Sigma-Aldrich Chemistry) and polystyrene (PS, from Bangs Laboratories, Inc.) spheres of 7 and 5 μm in diameter, respectively, as well as GaN nanowires nominally 10 μm in length and 300 nm in diameter [51,52]. All colloidal particles were dispersed in a CLC host either via mechanical mixing or solvent exchange, with both methods producing comparable quality of dilute colloidal dispersions [49–52].

Our LC-colloidal sample cells were fabricated from glass microscope slides cleaned in water and detergent, sonicated at 60 °C, and sequentially rinsed with acetone, methanol, and isopropyl alcohol. The slides were then dried and plasma cleaned. Planar alignment boundary conditions for n on the cleaned substrates were set by spin coating them with either an aqueous solution of polyvinyl alcohol (PVA, 1 wt%) at 8500 rpm or with polyimide PI-2555 at 3500 rpm, after which they are baked for a minimum of 1 h at 100 °C and 270 °C, respectively. These alignment layers were subsequently rubbed with a velvet cloth, which forced LC molecules to align along a rubbing direction. These substrates were assembled in a planar cell configuration with a director pretilt angle of 3°–6° [48,49] and were constructed with a cell gap thickness typically set using spherical spacers dispersed in a UV curable epoxy (NOA-61, Norland Products). The CLC colloidal dispersions were infused into these cells using capillary forces and the cells were sealed with a fast setting epoxy. Using HOT, we manipulated various colloidal species, including superparamagnetic and dielectric microspheres and semiconductor nanowires, as well as studied their interactions within the periodic lattice of layer deformations as described below.

# III. RESULTS

## A. Characterization of potential landscape controlled by electrically tunable undulations

The helicoidal ground-state structure of CLCs has been extensively studied [1]. A key feature of these soft matter systems is a directionally periodic structure characteristic of lamellar phases [41,44], with layers that tend to preserve equidistance equal to $p/2$ [1,41,44] [Figs. 1(a) and 1(e)] while easily bending in response to external stimuli and boundary conditions [1]. The three-dimensional morphology of layer configurations that satisfies these constraints provides a rich experimental platform controlling colloidal self-assembly that includes, for the purposes of this work, periodic distortions of the cholesteric lamellae, called "undulations," depicted in Figs. 1(b), 1(d), and 1(f) [38–43]. In general, undulations can occur in response to mechanical dilations, temperature change-induced variations of the cholesteric pitch, and applied electric and magnetic fields, with the latter being perhaps of the greatest interest for technological applications and for controlling particle assembly. When an electric field is applied normal to the cholesteric layers of a CLC material with a positive dielectric anisotropy, they tend to reorient parallel to the field. However, these reorientations are rotationally hindered by anchoring forces at the confining substrates of a LC cell, which results in periodic patterns of these undulating layers [Figs. 1(b), 1(d), and 1(f)], with a periodicity dependent on sample thickness and equilibrium pitch. The amplitude of these undulations can be tuned within a rather broad range by an applied voltage above a certain well-defined threshold [38–40,42,43]. Since undulation patterns emerge from a competition between the elastic, surface anchoring, and dielectric terms of free energy of a CLC with positive dielectric anisotropy [1,2], layer undulations are observable when the applied voltage reaches a certain critical threshold value $U_{th}$. This threshold voltage depends on the strength of surface anchoring at confining substrates and scales as $U_{th} \propto (d/p)^{1/2}$ for the limit of infinitely strong surface anchoring [1,2,38–40] and as $U_{th} \propto d/[p(d + 2\xi_e)]^{1/2}$ in softly anchored cells [42,43]. In the above expressions, $d$ is the cell thickness, and $\xi_e$ is the surface anchoring extrapolation length characterizing cholesteric anchoring at the confining surfaces, which is usually found to be on the order of the equilibrium CLC pitch p [42–44]. For the infinitely strong surface anchoring and material parameters of the studied CLCs (Table I), as well as cells with $p = 5$ μm and $d = 60$ μm, we can estimate that $U_{th} \approx$ 3.7 V for 5CB based cholesteric systems and $U_{th} \approx 9.9$ V for the ZLI-3412 based systems [38–40,42,43]. Our CLC cells have a large ($d/p > 5$) cell thickness to pitch ratio, and, thus, the relaxed undulations patterns are roughly square-periodic, with the two mutually orthogonal wave vectors characterizing undulations aligned parallel and perpendicular to the antiparallel rubbing directions at the confining substrates (Fig. 1).

Colloidal particles inserted into a ground-state CLC cause localized bending of cholesteric layers even at no applied fields [Figs. 2(a) and 2(c)]. When these distorted layers undergo undulations in response to an applied electric field [Figs. 1(d) and 1(f)], colloidal particles become elastically trapped within particular sites of a two-dimensional (2D) lattice of deformations where layer bending is the largest [Figs. 2(b), 2(d), and 2(e)]. This CLC-mediated localization of spherical particles assures that the overall free energy cost due to incorporation of particles and layer deformations prompted by both particles and applied fields is minimized. Each particle can be controllably trapped in layer deformations at different distances from the confining substrates [compare Figs. 2(d) and 2(e)], which allows for an additional control over the 3D localization of particles in the undulating cholesteric structure along the sample depth direction. Multiple particles embedded within the CLC are elastically trapped in separate deformation sites of the undulations lattice (Fig. 3) while the direct elastic interparticle interactions between the colloidal inclusions (present also at no fields) are hindered by potential barriers due to the interactions of individual particles and elastic distortions of the undulations lattice. To characterize the stiffness of this elastic trapping, we use video microscopy to record fluctuations of the in-plane position of a spherical particle trapped within a given deformation site, such as shown in Fig. 2(b), for example. Upon applying a voltage $U = 11$ V we measure and analyze the probability histogram $P(r)$ of the particle's lateral position r, which allows us to determine the trap's potential barrier preventing the escape of the particle from the layer-deformation-enabled trapping site [Fig. 4(a)]. This potential well is experimentally characterized by using video microscopy and subsequently inverting the Boltzmann relation $P(r) \propto \exp(-\Delta W_{pe}/k_B T)$, where $k_B$ is the Boltzmann constant, $T$ is the absolute room temperature, $\Delta W_{pe} = (1/2)k_{et}r^2$ is the elastic potential energy required to move the particle by a distance r from its equilibrium location (here we assume a parabolic potential for the near-equilibrium positions of the particle in the trapping site), and $k_{et} \approx 0.27$ pN/μm is a stiffness of the elastic trap at room temperature obtained from fitting experimental data [Fig. 4(a)].

We further study the physical underpinnings of how different particles become entrapped in an undulation deformation site produced by applying a voltage to the sample [Figs. 2(b), 2(d), 2(e), 3(b), and 3(d)]. To measure the elastic trapping force exerted on the particle by the deformation site, we use HOT to optically reposition it from its initial trapped position to a location between neighboring trapping sites, as marked by a yellow cross in the inset of Fig. 4(b). When the particle is released from the HOT trap, it experiences attraction towards the center of the layer deformation trapping site. We analyze the resulting video microscopy frames and measure a rapidly decreasing center-to-center separation $s$ between the particle center and its final equilibrium location within the elastic trapping site of the undulations pattern [Fig. 4(b)]. Since this trapping force is balanced only by a viscous drag force in the regime of low Reynolds and Ericksen numbers which characterize our system [5,25], inertial and elastic-viscous coupling effects may be neglected. Therefore, results of these experiments allow us to measure the trapping force $F_{tr}$ using the time dependent separation $s(t)$, which is determined as $F_{tr} \approx c_f (ds/dt)$, where $c_f \approx 1.42$ Ns/m is a friction coefficient measured for a spherical particle undergoing Brownian motion [25] in a ground-state CLC at no fields [50]. Within the interaction distances studied, the elastic trapping of a spherical particle by a given deformation site shows a Hookean-like behavior with a maximum force on the order of 1 pN [Fig. 4(c)], with a corresponding interaction potential difference on the order of hundreds of $k_B T$ [Fig. 4(d)], and is dependent on the applied voltage $U_{th}$. The stiffness of the elastic trap extracted from a linear fit $F_{tr} = k_{et}s$ of experimentally measured $F_{tr}$ [Fig. 4(c)] at $U = 11$ V is found to be $k_{et} \approx 0.26$ pN/μm, which is comparable to that extracted from above described experiments characterizing Brownian motion of a particle in the elastic trap [Fig. 4(a)]. The elastic interactions between the occupied deformation site and a nearby particle are still attractive. Depending on the initial spatial position of colloidal particles with respect to each other, several particles can be entrapped in one deformation site in one spot or separated by cholesteric layers at different distances from the substrates. Characterization of interactions between colloidal microspheres and voltage controlled periodic patterns of undulations shows that a CLC colloidal system is highly reconfigurable and that low-voltage fields allow us to induce a periodic voltage-dependent potential landscape amenable to controlling the 3D positions of colloidal particles.

**B. Control of the onset of undulations by particle arrays and control of self-assembly, self-alignment, and dynamics of particles by undulations**

The previous section's findings/observations regarding particle-CLC interactions mediated by the LC elasticity and formation of periodic undulations can be extended from individual particles and pairs of particles to many-body interactions involving multiple colloidal objects. We observed that the presence of particles influenced the onset of the undulations instabilities and that, in turn, periodic undulations further enrich elastic interactions between the CLC-dispersed colloidal particles. To show this, five melamine resin beads of 7 μm nominal diameter were positioned into an initial configuration shown in Fig. 5(a) using HOT and then released from the laser traps. These particles induced local perturbations of the CLC director field which are visible in polarizing optical micrographs [Fig. 5(a)]. Subsequently, periodic distortions were induced by gradually increasing the amplitude of a 1-kHz sinusoidal ac electric field to a final peak-to-peak value of 6.5 V applied in the direction normal to the CLC lamellar plane (and the plane of the cell). We find that this induces layer distortions which first nucleate around the HOT-configured arrangement of colloidal particles at a voltage ~0.5 V below the critical threshold voltage $U_{th}$ needed to induce undulations without colloidal inclusions in the same sample away from particles. As an intermediate step, the layer's distortions then form a metastable undulation pattern which is locally predefined by initial HOT-configured arrangement [Fig. 5(b)]. Over a time period of approximately 30 s, the layer's undulations finally relax to a ground-state lattice configuration, which remains stable in terms of lattice site position, lattice dimensions, and lattice orientation unless disturbed by encroaching topological defect structures, such as Lehmann clusters, or the so called "oily streaks" [1,2]. Each undulation pattern that develops within the sample at fields above the critical field ($U > U_{th}$) locally reflects the initial locations of the particles in its vicinity. The eventual slow (seconds to tens of seconds) relaxation of the undulation pattern forces the colloidal assembly to reorder in a periodic manner, as dictated by the free energy minimization driving the system towards the ground-state lattice of undulations that we discussed above. If we then turn the applied voltage off or reduce it to values well below the $U_{th}$ threshold for the periodic distortions, the colloidal particles are found in a new positional arrangement [Fig. 5(c)], which typically significantly differs from their initial configuration defined by the laser tweezers [Fig. 5(a)]. This repositioning of particles is caused by transient effects related to relaxation of the lattice of undulations upon turning voltage off. We repeat the above-described sequence of releasing particles from traps, inducing undulations by applied voltage and then turning the voltage off for an ensemble of SPMBs arranged in various initial configurations [Figs. 5(d)–5(l)]. We find that by selectively positioning the SPMBs, we can reproducibly dictate the local orientation and spacing of the induced undulations lattice at the onset of its formation (Fig. 5). The undulations, in turn, alter positions of the colloidal particles released from the laser traps of HOT as the undulations pattern slowly relaxes to eventually become square-periodic in nature, with the lattice wave vectors eventually aligning either parallel or perpendicular to the cell's substrate rubbing direction [Fig. 5(k)]. An initial arrangement of particles often also induces defects in the square-periodic pattern of undulations, such as those with edge dislocations of Burgers vector equal to the undulations lattice period shown in Fig. 5(b), albeit these nonequilibrium structures also tend to relax to defect-free patterns with time. The spatial locations of particles at applied fields, collocated within the regions of the strongest distortions in the cholesteric layer patterns, do not persist upon turning off the applied voltage [Figs. 5(c), 5(f), 5(i), and 5(l)]. Upon relaxation of undulations, these particles undergo free diffusion within the CLC, as well as resume exhibiting elastic interactions characteristic of such colloidal particles in a system of flat cholesteric lamellae. In response to an instantaneous application or reduction of the applied electric field, the colloidal particles are found collectively repositioning within the 3D volume of the LC cell, including along the sample's depth, as indicated by the image defocusing apparent from the comparison of Figs. 5(h) and 5(i). This kinetics of particles is associated with strong transient changes of the cholesteric layer patterns that transform between the two equilibrium layer configurations at the applied field and without it.

The spherical symmetry of SPMBs and melamine resin microparticles allows us to probe their interactions with voltage-tunable CLC undulations that alter only the positional or spatial arrangements of these colloidal objects. To demonstrate the possibility of additionally controlling the orientational degrees of freedom of colloidal inclusions, we use shape-anisotropic colloidal particles, such as GaN nanowires (Fig. 6) with tangential anchoring conditions at their surfaces that were established previously [51,52].We prepared dilute dispersions of GaN nanowires of length $L = 10$ μm and effective diameter $D = 300$ nm [the inset of Fig. 6(b)]. We then manipulated these dispersed high-aspect-ratio GaN nanowires to localize in specified locations in the midplane of the CLC cell using HOT [52] and subsequently induced a periodic undulation lattice (Fig. 6). Similar to melamine resin and SPMB colloidal particles, these GaN nanowires served as nucleation centers for the undulations lattice, prompting a reduction in the threshold voltage $U_{th}$ required for observing the onset of undulations in the vicinity of particles by approximately 0.2 V relative to the value of $U_{th}$ established for the sample away from these colloidal inclusions. GaN nanowires have a tangential anchoring at their surface, which makes them follow the local director and be confined within the local cholesteric layers [52]. When cholesteric layers are periodically bent into an undulation pattern at applied fields above $U_{th}$ [Fig. 1(f)], the nanowire inclusions rotate while being effectively mechanically coupled to the layer orientations, tilting out of the plane of the cell and staying roughly orthogonal to the local helical axis $\chi(\mathbf{r})$ (Fig. 6). These $U$-dependent layer tilt angles are observed to reach up to 30° out of plane (Fig. 6). The elastic interactions of the CLC suspended nanowires with the periodic lattice of undulations are further enriched by the dielectrophoretic effects that originate both from the shape-anisotropic nanowires having dielectric properties rather different from that of the CLC, as well as from the modulations of the effective dielectric constant within the CLC cell bulk caused by periodic tilting of $\chi(\mathbf{r})$ in response to the applied field. Further study is needed to quantify the contribution of the elastic and dielectrophoretic effects, albeit we posit that elastic effects dominate for these colloids as we find no observable nanowire tilting at applied voltages well below the threshold value. This may indicate that the undulations-induced rotation of GaN nanowires originates mainly from the elastic interactions that make the nanowires align with respect to the local director or layer orientations, similar to what we observed for such GaN nanowires following cholesteric layers around edge dislocations [52].

Interestingly, a multistable but well-defined alignment of a GaN nanowire with respect to an undulations lattice is observed to occur following each of the repeated undulation nucleations, as shown in Fig. 6(a). Each time undulations were prompted to dissipate by turning off the applied voltage, we either manipulated GaN nanowires using HOT to a new location, or let thermal motion and gravity align the nanowire along a new direction. Upon inducing a new undulations lattice, the nanowire aligns along a new direction out of several possible multistable states [Fig. 6(b)]. Likewise, observing an ensemble of nanowires within a given field of view in a microscope with multiple undulation trapping sites shows the same multistability of nanowire orientations [Fig. 6(a)]. Summarizing these observations, eight well-defined observed metastable orientations of the nanowires relative to the square periodic lattice of undulations are shown in Fig. 6(b), as indicated using red arrows. The alignment of rodlike shape-anisotropic particles dispersed in the ground-state structures of nematic LCs was previously demonstrated for nanoparticle orientations parallel and perpendicular to the director [5,7,9,12,22], thus providing only a limited spectrum of accessible types of control of orientational degrees of freedom of these anisotropic colloidal particles. In the ground-state CLCs, these types of alignments are enriched by the helicoidal structure of the director field [51,52], with each anisotropic nanoparticle locally exhibiting behavior similar to that characteristic of nematic dispersions. Our current studies show that these LC-colloidal interactions can be dramatically altered by the patterns of undulations that highly enrich them and yield the eight-stable orientational configurations with tilted GaN inclusions (Fig. 6).

### C. Control of nonequilibrium dynamics of particles by undulations

Dynamic, noncontact manipulation of colloidal particles in CLC systems has provided insights into the structure, microrheology, generation, and control of topological defects in soft matter [46,48–52]. Cholesteric layer undulations can further enrich the capabilities of magnetic and optical control of colloidal particles in CLCs. Our present study demonstrates that, in addition to controlling spatial localization, self-assembly, and self-alignment of colloidal inclusions at the equilibrium, as discussed above, CLC undulations also provide a means to alter nonequilibrium complex dynamics of particles. We show this by employing holonomic optical and magnetic control of SPMB particles in conjunction with controlling the director field configurations of CLCs using electric fields [48–50]. Our previous studies demonstrated that anchoring-mediated mechanical coupling between colloidal inclusions and the local helicoidal director structure of CLCs allows for translating these particles in well-defined directions across the cholesteric layers via rotating them around this helical axis optically [46] or magnetically [48,50]. In the present work, we demonstrate the possibility of more complex well-controlled 3D spatial translations of magnetically rotated SPMB colloidal particles as a result of their strong coupling to the spatial patterns of a distorted cholesteric structure with undulations. To illustrate these capabilities, a periodic undulations lattice was produced in a ∼60-µm-thick CLC cell filled with a left-handed mixture of 5CB and cholesteryl pelargonate wherein SPMBs were magnetically rotated around an axis normal to the cell substrates at 0.25 Hz using a 40-G ac magnetic field (Fig. 7). The ensuing dynamics of colloidal particles showed a behavior different from that previously observed in CLCs with a ground-state uniform helicoidal structure [46,48–50]. We tracked specific periodic spatial locations of sites within the square-periodic undulations lattice using video microscopy and observed them to shift slightly with time [Fig. 7(a)] due to thermal fluctuations, relaxation of dislocation defects in the lattice, and other transient effects [1,2,44]. SPMB motion was tracked with respect to time during the magnetic field rotation in the *x-y* plane (around the axis normal to cell substrates). POM micrographs, along with 3D nonlinear optical images such as the ones shown in Fig. 7(b), allow us to distinguish between the 2D lattice sites corresponding to the strong distortions of layers at the top and bottom confining surfaces. A SPMB initially located at the bottom confining surface at position 1 [Figs. 7(b) and 7(c)] was magnetically rotated in a clockwise (CW) direction, which, due to the mechanical coupling between the local helicoidal director and the colloid surface through anchoring, translates the SPMB along the local helical axis to the proximity of the top substrate. Because the chiral dopant and the ensuing helicoidal structure of CLC were left-handed, the CW rotation of the field resulted in a translation of the colloidal particle from the bottom confining surface to the top substrate, ending at position 2 [Figs. 7(b) and 7(d)]. A subsequent counterclockwise (CCW) rotation of the colloidal particle moved the SPMB from the top confining surface back to the bottom one at position 3 [Figs. 7(b) and 7(e)]. Importantly, the particles did not move directly upward or downward in straight paths, but also shift in the lateral directions as compared to their initial positions (for example, compare positions 1 and 2 along the *x* axis in Figs. 7(b)–7(d). These lateral translations are much larger than the small displacements of lattice sites tracked in Fig. 7(a). This kinetic behavior is mediated by the entrapment of the colloidal inclusions in the regions with the strongest layer deformations, which have complex morphology and are 3D in nature. Generally, due to the symmetry of the undulations lattice, there are eight possible pathways between the top and bottom confining surfaces for each initial position in the lattice, all of them yielding a well-defined lateral translation corresponding to the periodicity of the undulations pattern, four of them along and four of them diagonal to its wave vectors [Fig. 7(e)]. Due to the nature of the potential landscape introduced by cholesteric undulations [Figs. 2(b) and 2(e)], each path between confining surfaces is, in principle, equally likely, although the director pinning at particle surfaces and the ensuing surface anchoring memory effects may lift this degeneracy. By repeating these manipulations many times, we have observed that a variety of factors (e.g., local motion of lattice sites due to relaxation of edge dislocations within the periodic lattice, a size of the particle with respect to a thickness of the cholesteric layer and its surface anchoring) can alter these processes. For example, as shown in Fig. 7(e), we observed the particle following a path (1-2-4) where a position 2 is not at the top as in Figs. 7(b)–7(d) but in the middle of the cell and a position 4 is at the top substrate. In addition, fine tuning of the details of magnetic rotation and additional manipulations with laser tweezers allows for a robust and repeatable selection of one out of eight possible pathways of motion within the cell volume from surface, with a resulting translation within the lateral plane by a half of the periodic lattice period, as illustrated in Fig. 7. Beyond this magnetically induced rotation, similar types of complex dynamics of particles within the CLCs with voltage-tunable undulations can be potentially also induced by gravity and electrophoretic effects [53].

## IV. DISCUSSION

Although cholesteric blue phases provide a 3D periodic potential landscape for trapping and assembly of colloidal inclusions, these lattices can have only small submicron range of lattice periods, typically on the order of hundreds of nanometers [1,54]. In principle, the potential landscape explored by colloidal inclusions in CLCs can be pre-engineered [1] by controlling undulation patterns through varying $p$, $d$, the elastic and dielectric properties of the LC host, the strength of surface anchoring at confining cell substrates, etc. The period of the square-periodic lattice of undulations is $\lambda_u = C(6K_{33}/K_{22})^{1/4}[p(d + 2\xi_e)]^{1/2}$, where $C$ is a constant of the order of unity that depends on surface anchoring, and $K_{22}$, $K_{33}$ are the twist and bend elastic constants, respectively [42,43]. By controlling $p$ over a range from ~100 nm to hundreds of microns and $d$ from ~1 μm to hundreds of microns, $\lambda_u$ can be varied from the submicron range to the range of hundreds of micrometers by adjusting the cell and material parameters, as well as varying the surface treatment that controls the strength of surface anchoring (and $\xi_e$). This establishes CLC undulations as a versatile tool for the patterning of colloidal inclusions that expands similar capabilities provided by the ground-state blue phases of CLCs [54] as well as different types of topographic features at confining surfaces of LC cells [27].

The patterns of CLC undulations correspond to a complex 3D spatial modulation of director field that introduces the corresponding periodic modulations of optical phase retardation and the effective refractive index [1]. However, the ability of controlling optical properties of the electrically reconfigurable CLC samples is often limited by the contrast of the effective refractive index, which, in turn, is constrained by the intrinsic optical anisotropy of the nematic LC host used to form CLCs (typically ≈0.2 or less for commercially available nematic materials such as 5CB and ZLI-3412 used in our experiments). One potential strategy for overcoming limitations due to these material-specific constrains involves doping CLCs with semiconductor and metallic colloidal particles, which can alter the light-matter interactions to significantly enrich the physical behavior and properties of the ensuing composites. Our observation that the colloidal inclusions tend to spatially localize in the regions of an undulations lattice associated with strong layer deformations indicate that particles inserted into such electrically controlled CLCs can enhance the ability of colloidal patterning, as well as the CLC colloidal composite's effective refractive index modulation and other optical properties. The use of plasmonic and luminescent nanoparticles in conjunction with undulation-enabled patterning may allow for the formation of reconfigurable active and passive metamaterials with a host of potential technological applications.

The elasticity-mediated interactions and surface anchoring-mediated alignment of colloidal inclusions in CLCs with tunable undulations patterns can be further enriched by dielectrophoretic effects (and also electrophoretic effects when these particles are charged), as we already briefly mentioned above. Even in isotropic fluid hosts, colloidal particles with dielectric constants different from that of the surrounding medium are known to strongly interact with the gradients of electric field (e.g., when these gradients are produced by using patterned electrodes) [55–57]. Shape-anisotropic or composition-anisotropic particles in the isotropic fluid hosts also rotate and align with respect to the applied electric field direction in order to minimize the dielectric term of the free energy [55–57]. In LC host fluids with large values of dielectric anisotropy, including CLCs, these effects are dramatically enriched by the fact that the patterns of director field are also the patterns of the effective dielectric constant variations. The ensuing gradients of electric field arising from these dielectric constant patterns can have both the aligning and spatially localizing (entrapping) effects on colloidal inclusions with different shapes and compositions, tending to minimize the dielectric terms of the free energy by positioning particles in different locations within the 3D volume of the sample as well as aligning them with respect to the wave vectors of corresponding periodic patterns. The dielectrophoretic effects can be synergistic or compete with the elastic and surface-anchoring-based forces and torques that we discussed above, further enriching the abilities of controlling colloidal inclusions in CLCs, albeit the complete understanding of this complex interplay will require numerical modeling and will be pursued and discussed elsewhere.

## V. CONCLUSIONS

We studied the interaction between colloidal particles and layer undulations induced in a cholesteric liquid crystal under an ac electric field applied normal to the cholesteric layers. We have shown that the inclusion of colloids reduces the threshold voltage for the local appearance of undulation instability and allows for controlling the local structure of undulation lattices at the onset of this instability. Elastic coupling between the colloidal particles and a CLC with voltage-induced undulations allows for controlling 3D locations of colloidal particles within the sample volume, their assembly into arrays mimicking the underlying undulations patterns, as well as multistable alignment of shape-anisotropic colloids such as GaN nanowires. Moreover, voltage-tunable potential landscapes formed due to layer undulations allow for robust control of complex particle dynamics that are not otherwise accessible in the ground-state LC systems at zero applied fields. These LC-colloidal interaction modes lend insight into the rich physical phenomena associated with CLC-colloid elastic interactions, as well as point the way towards reconfigurable, guided self-assembly of tunable colloidal composites with potential applications in diffraction optics and photonics. Our findings can be further extended to LC colloidal systems with optical patterning of LC alignment at confined surfaces and a broad range of active colloidal inclusions [53,58–60] driven by local energy conversion or external fields.

## ACKNOWLEDGMENTS

We thank Taewoo Lee for discussions and assistance with nonlinear optical imaging and Kris Bertness for providing GaN nanowires [image in the inset of Fig. 6(b)] [46,51,52]. We acknowledge support of National Science Foundation Grant No. DMR-1410735.

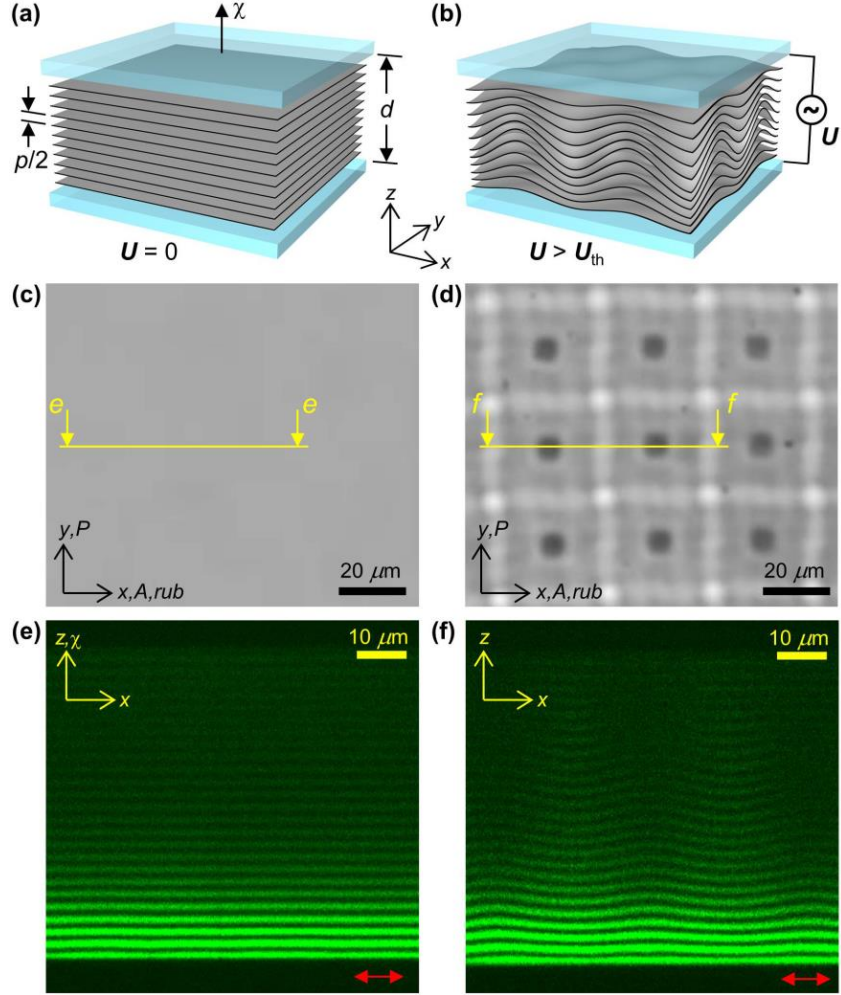

FIG. 1. Periodic layer undulations in a cholesteric liquid crystal (ZLI-3412 doped with CB15) in a thick planar cell ($d \approx 62$ μm). (a) Schematic diagram of cholesteric layers of thickness $p/2$ represented by gray planes and confined between two glass substrates with ITO electrodes and alignment films. The arrow marked with "$\chi$" shows the direction of a helical axis normal to the confining substrates. (b) Schematic representation of periodic distortions of cholesteric layers under voltages $U > U_{\text{th}}$. (c),(d) POM and (e),(f) 3PEF-PM vertical cross-section textures of uniform (c),(e) and undulating under an applied ac voltage $U \approx 10.2$ V at 10 kHz (d),(f) cholesteric layers. Vertical cross-section textures (e),(f) were taken along yellow lines in (c),(d) marked, respectively, by letters "$e$" in (c) and "$f$" in (d). "$P$" and "$A$" in (c),(d) indicate the crossed linear polarizer and analyzer, with the orientations shown using arrows. The direction of rubbing at the alignment layers is marked by "rub." The polarization direction of the 3PEF-PM excitation light in (e),(f) is shown using red double arrows.

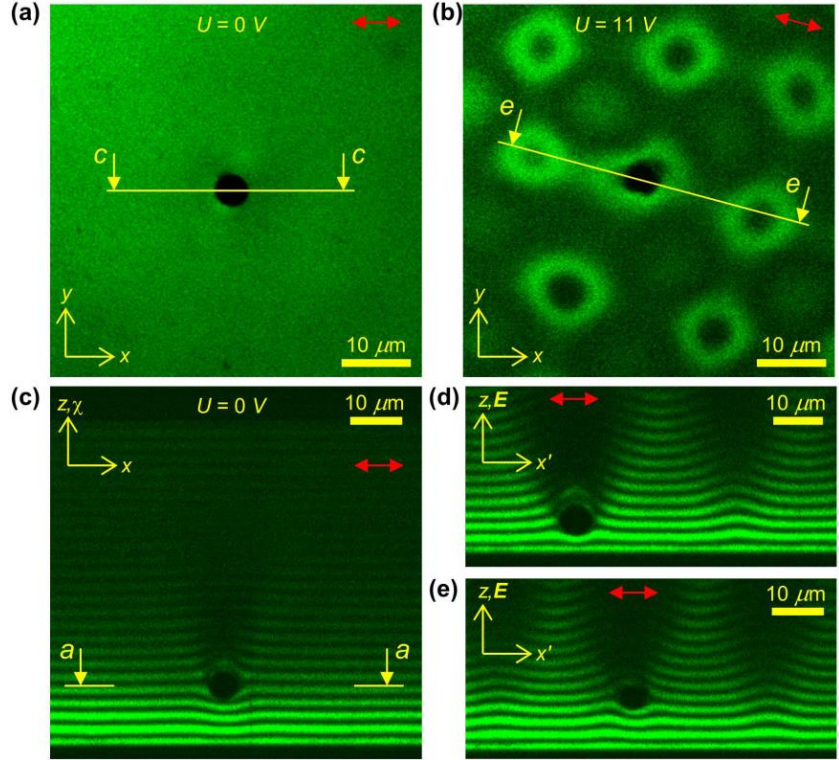

FIG. 2. In-plane (a),(b) and vertical cross-sectional (c)–(e) 3PEFPM images of cholesteric layers in a sample of ZLI-3412 doped with CB15 and with an embedded colloidal PS particle. (a),(c) The colloidal PS particle in the homogeneous cholesteric layered structure. (b),(d),(e) PS particle trapped in deformations of undulating cholesteric layers under an applied electric field $E$ (the applied voltage was $U = 11$ V). Cross-section textures (c),(e) were taken along yellow lines in (a),(b) marked respectively by letters "$c$" in (a) and "$e$" in (b). Horizontal yellow lines in (c) marked by downward arrows and a letter "$a$" indicate a $z$ level of the in-plane $x$-$y$ 3PEF-PM image shown in (a).

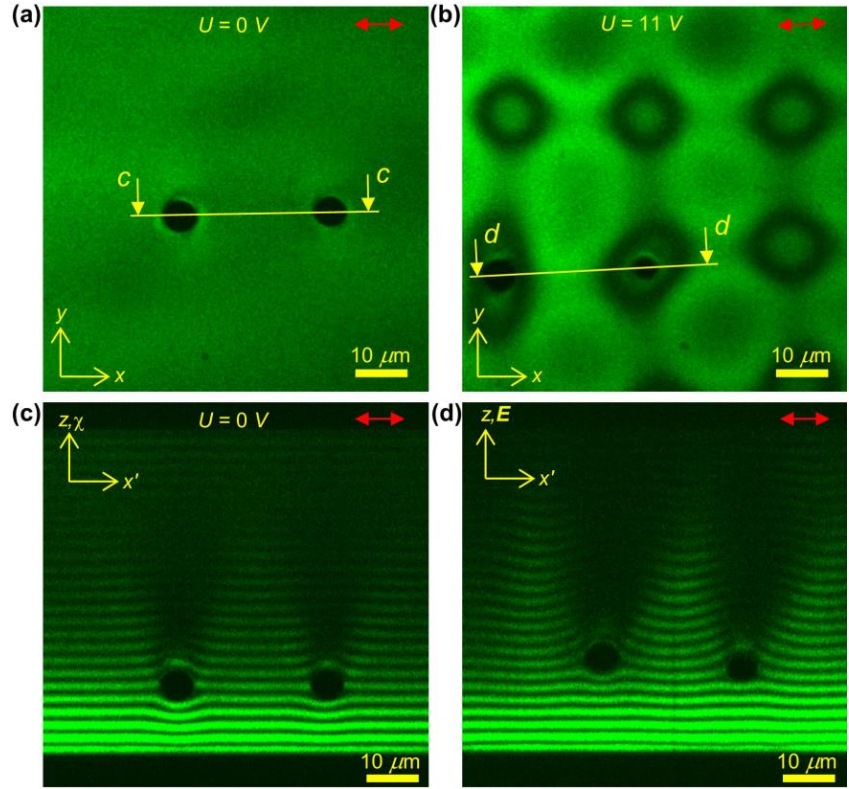

FIG. 3. In-plane (a),(b) and vertical cross-sectional (c),(d) 3PEFPM images of an assembly of particles in a deformations landscape of cholesteric layers undulations. (a),(c) A pair of colloidal PS particles in the homogeneous cholesteric structure (ZLI-3412 doped with CB15). (b),(d) The same colloidal particles trapped in deformation sites of undulating cholesteric layers under an applied electric field E (the applied voltage was $U = 11$ V). The 3PEF-PM vertical cross-section images (c),(d) were taken along yellow lines in (a),(b) marked by downward arrows and, respectively, letters "c" in (a) and "d" in (b).

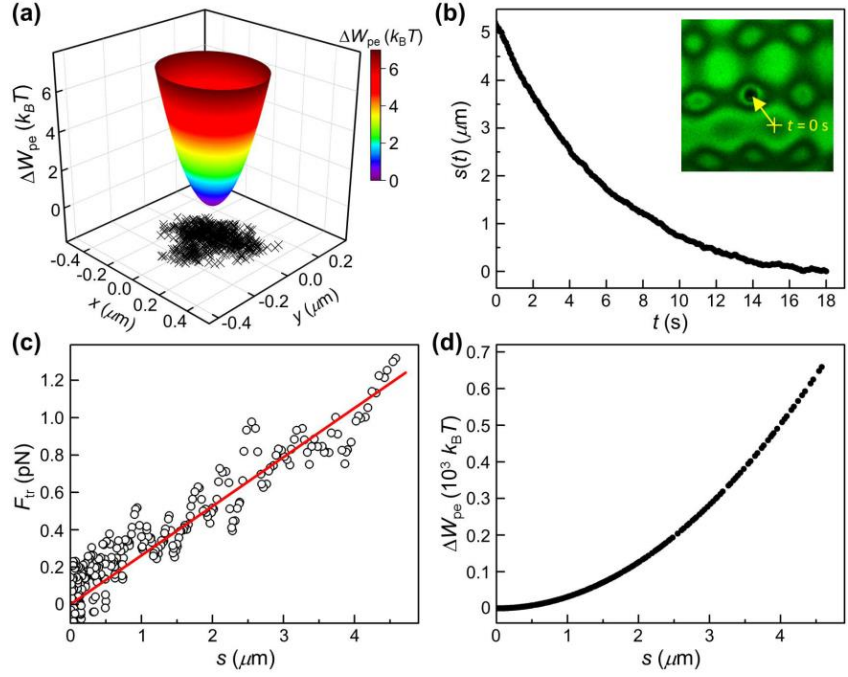

FIG. 4. Characterization of the elastic trapping of colloidal particles in a deformation landscape of cholesteric LC (ZLI-3412 doped with CB15) layers undulations. (a) Positional fluctuations of a colloidal PS particle trapped in a deformation site and its corresponding potential well; a surface plot of the potential well is a Gaussian fit to experimental data. (b) Time dependence of a center-to-center separation distance $s$ between a PS particle and a deformation site at the applied voltage $U = 11$ V; the inset shows the initial position (a yellow cross) of a particle outside of the deformation site and its final trapped position within the deformation site. (c) Separation dependence of the trapping elastic force exerted on a PS particle by the deformation site; the red line is a linear fit to the experimental data (open circles). (d) Elastic trapping potential vs the center-to-center separation between a PS particle and a deformation site as calculated using $k_{et} \approx 0.26$ pN/μm extracted from a linear fit in (c).

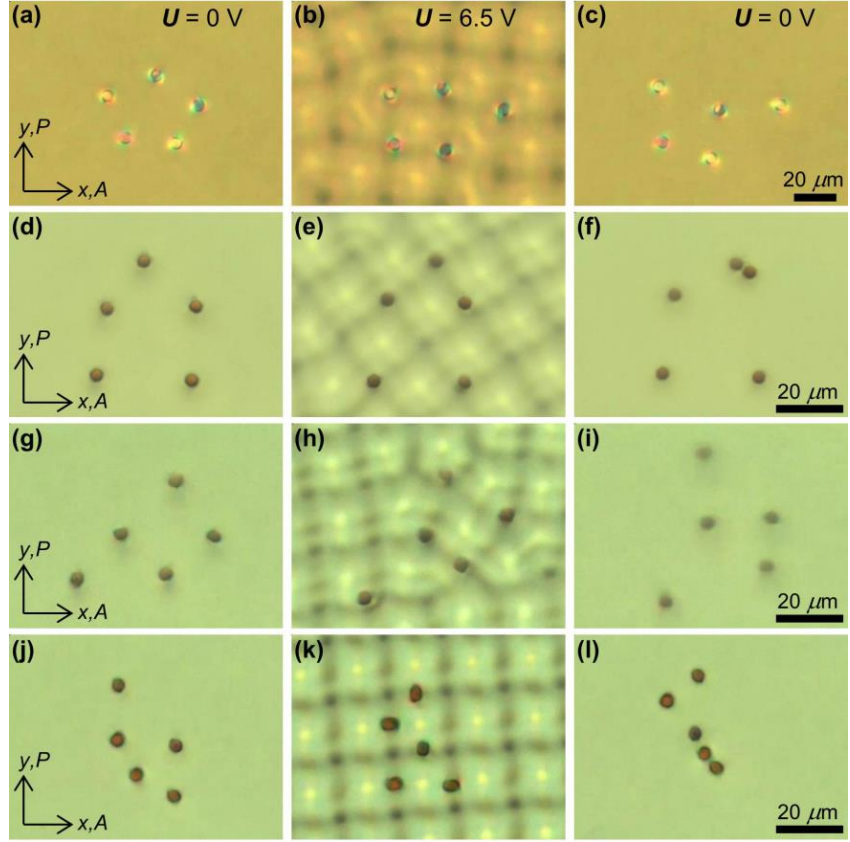

FIG. 5. Interaction of colloidal particles with periodic layer undulations in a CLC (5CB doped with CB15). (a) Melamine resin beads of 7 μm nominal diameter manipulated via HOT to form an initial configuration. (b) Periodic distortions are induced by applying a voltage, which first nucleates in the vicinity of the colloidal particle positions due to particle-induced deformations of the ground-state cholesteric structure. Over a period of ≈30 s, these cholesteric layer distortions relax to their ground-state square-periodic configuration without edge dislocations in the lattice (such as the one prompted by particle positions visible in the micrograph). Each transient lattice dislocation has a Burgers vector equal to one lattice period. (c) After the electric field is turned off, each colloidal particle position is influenced by transient layer realignment associated with the relaxation of undulations after which the particles undergo random translational diffusion. (d)–(l) SPMBs of 4.5 μm nominal diameter are manipulated using a combination of HOT and voltage-induced undulations in a similar manner as in (a)–(c) by defining the initial arrangement (d),(g),(j), applying voltage to induce undulations (e),(h),(k), and then removing the voltage (f),(i),(l).

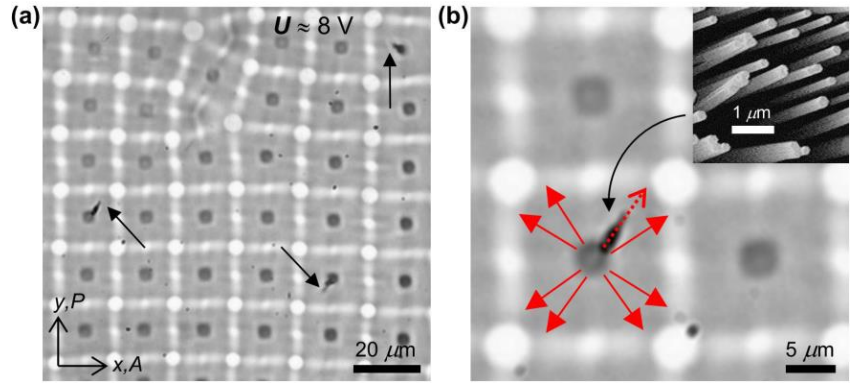

FIG. 6. CLC undulations lattice induced multistable alignment of GaN nanowires dispersed in 5CB doped with CB15. (a) POM image showing the alignment of a GaN nanowire upon formation of a periodic undulations lattice. Individual GaN nanowires are indicated by black arrows. (b) Repeated observations of multiple particles indicate that GaN nanowires tend to align along one of eight different "multistable" orientations with respect to the undulations lattice, and are tilted out of the lateral plane of the CLC cell by 20–30°, depending on their depth locations within the bulk. Selection of orientations among the eight "stable" directions (the one seen in the micrograph and marked by a dashed red arrow and the seven more possible shown using red arrows) is nearly random, with a new orientation typically occurring each time an undulation lattice is induced. The inset in (b) shows a scanning electron microscopy image of a "forest" of GaN nanowires growing from a substrate [46,51].

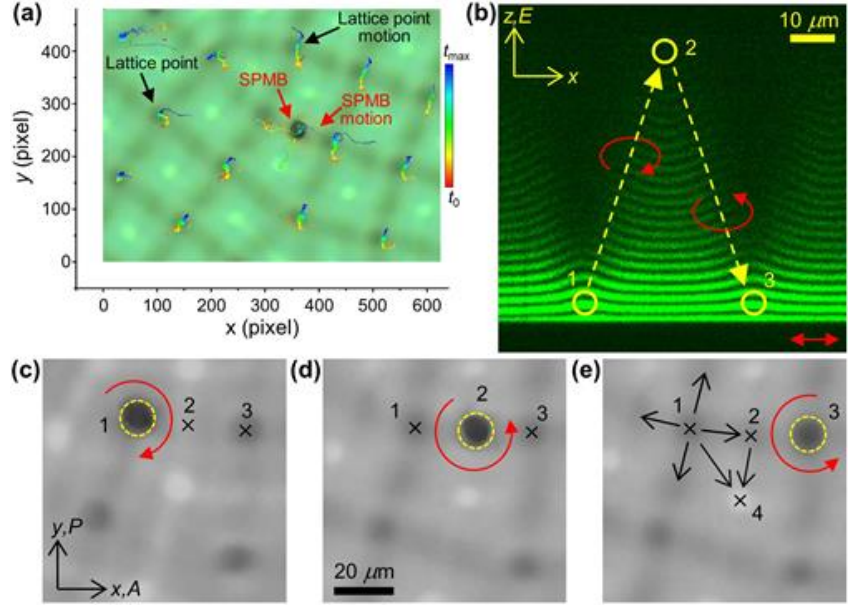

FIG. 7. 3D spatial manipulation of SPMBs induced by magnetic rotation and mediated by CLC undulations in 5CB doped with cholesteryl pelargonate. (a) A polarizing optical micrograph of an undulations lattice showing movement of the lattice points and motion of the SPMB in the lateral directions of the CLC cell when a 40-Gs magnetic field is rotated in the x-y plane at a frequency of 0.25 Hz in both CW and CCW directions. (b) 3PEF-PM vertical cross-sectional image obtained along the line 1-2-3 in (c)–(e), which is used to schematically show the translational motion of SPMB in the vertical $x$-$z$ plane. Red circular arrows show the sense of the in-plane rotation of a magnetic field and SPMB. Yellow dashed arrows show the corresponding translational motion of a SPMB across cholesteric layers. (c),(d) POM textures showing rotational and corresponding in-plane $x$-$y$ motion of SPMB from position 1 to positions 2 and 3 (marked by a cross) caused by magnetic CW rotation to translate from position 1 to 2 (c),(d) and CCW rotation to translate between positions 2 and 3. (d),(e) The final positions of a SPMB corresponding to locations 1, 2, and 3 are marked by yellow circles in (b)–(e). Black arrows in (e) show multiple in-plane projections of the trajectories that the SPMB initially positioned at location 1 can take upon rotation by a magnetic field.

Table 1. Material parameters of used nematic LC hosts and helical twisting power of chiral additives when doped into these nematic hosts: $K_{11}$, $K_{22}$, $K_{33}$ are Frank elastic constants, $\Delta n$ is birefringence (optical anisotropy), $\Delta\varepsilon$ is dielectric anisotropy, and HTP is the helical twisting power of a chiral dopant.

| Nematic LC host | $K_{11}$, pN | $K_{22}$, pN | $K_{33}$, pN | $\Delta n$ | $\Delta\varepsilon$ | HTP of cholesteryl pelargonate, $\mu$m$^{-1}$ | HTP of CB15, $\mu$m$^{-1}$ |
|---|---|---|---|---|---|---|---|
| 5CB | 6.4 | 3.0 | 10 | 0.175 | 13 | -6.25 | 7.3 |
| ZLI-3412 | 14.1 | 6.7 | 15.5 | 0.078 | 3.4 | not used | 6.3 |